\documentclass[aps,prd,preprint,tightenlines,superscriptaddress,floatfix,
  showpacs,byrevtex]{revtex4}
\usepackage{graphicx} 
\usepackage{dcolumn}  
\usepackage{amsmath}
\usepackage[sort&compress]{natbib}

\graphicspath{{ps}}

\def\ulnu{\ensuremath{u \ell \nu}}
\def\btoulnu{\ensuremath{b \to u \ell \nu}}
\def\btoclnu{\ensuremath{b \to c \ell \nu}}
\def\bpiplnu{\ensuremath{B\rightarrow\pi^+\ell\nu}}
\def\bpizlnu{\ensuremath{B\rightarrow\pi^0\ell\nu}}
\def\brhoplnu{\ensuremath{B\rightarrow\rho^+\ell\nu}}
\def\brhozlnu{\ensuremath{B\rightarrow\rho^0\ell\nu}}

\def\bomegalnu{\ensuremath{B\rightarrow\omega\ell\nu}}

\def\bdplnu{\ensuremath{B\rightarrow D^{+}\ell\nu}}
\def\bdzlnu{\ensuremath{B\rightarrow D^{0}\ell\nu}}
\def\bdsplnu{\ensuremath{B\rightarrow D^{*+}\ell\nu}}
\def\bdszlnu{\ensuremath{B\rightarrow D^{*0}\ell\nu}}
\def\bxulnu{\ensuremath{B\rightarrow X_u\ell\nu}}
\def\bxclnu{\ensuremath{B\rightarrow X_c\ell\nu}}
\def\bpilnu{\ensuremath{B\rightarrow \pi\ell\nu}}

\def\bdstarlnu{\ensuremath{B\rightarrow D^*\ell\nu}}
\def\bdlnu{\ensuremath{B\rightarrow D\ell\nu}}
\def\bddstarlnu{\ensuremath{B\rightarrow D^{**}\ell\nu}}
\def\bb{\ensuremath{B\overline{B}}\ }

\def\qsq{\ensuremath{q^2}}

\def\GeV{\ensuremath{\mathrm{GeV}}}
\def\GeVGeV{\ensuremath{\mathrm{GeV}^2}}

\def\mm2{\ensuremath{M^2_\mathrm{miss}}}
\def\deltae{\ensuremath{\Delta E}}
\def\mbc{\ensuremath{M_\mathrm{bc}}}
\def\ebeam{\ensuremath{E_\mathrm{beam}}}
\def\pb{\ensuremath{P_B}}
\def\btag{\ensuremath{B_\mathrm{tag}}}
\def\bzerobar{\ensuremath{\overline{B}{}^0}}

\def\pipeffglobal{\ensuremath{0.0401 \pm 0.0012}}
\def\pizeffglobal{\ensuremath{0.0581 \pm 0.0020}}
\def\rhopeffglobal{\ensuremath{0.0196 \pm 0.0006}}
\def\rhozeffglobal{\ensuremath{0.0339 \pm 0.0010}}
\def\omegaeffglobal{\ensuremath{0.0172 \pm 0.0007}}

\def\pipresultqsqafit{  \ensuremath{0.43 \pm 0.11}}
\def\pizresultqsqafit{  \ensuremath{0.26 \pm 0.09}}
\def\rhopresultqsqafit{ \ensuremath{0.74 \pm 0.29}}
\def\rhozresultqsqafit{ \ensuremath{0.72 \pm 0.15}}
\def\omegaresultqsqafit{\ensuremath{0.23 \pm 0.17}}

\def\pipresultqsqbfit{  \ensuremath{0.42 \pm 0.11}}
\def\pizresultqsqbfit{  \ensuremath{0.17 \pm 0.05}}
\def\rhopresultqsqbfit{ \ensuremath{1.01 \pm 0.28}}
\def\rhozresultqsqbfit{ \ensuremath{0.70 \pm 0.13}}
\def\omegaresultqsqbfit{\ensuremath{0.64 \pm 0.21}}

\def\pipresultqsqcfit{  \ensuremath{0.26 \pm 0.08}}
\def\pizresultqsqcfit{  \ensuremath{0.22 \pm 0.06}}
\def\rhopresultqsqcfit{ \ensuremath{0.81 \pm 0.21}}
\def\rhozresultqsqcfit{ \ensuremath{0.39 \pm 0.11}}
\def\omegaresultqsqcfit{\ensuremath{0.32 \pm 0.17}}
\def\piresultqsqccomb{  \ensuremath{0.31 \pm 0.07}}

\def\pipresultqsqsumfit{  \ensuremath{1.12 \pm 0.18}}
\def\pizresultqsqsumfit{  \ensuremath{0.66 \pm 0.12}}
\def\rhopresultqsqsumfit{ \ensuremath{2.56 \pm 0.46}}
\def\rhozresultqsqsumfit{ \ensuremath{1.80 \pm 0.23}}
\def\omegaresultqsqsumfit{\ensuremath{1.19 \pm 0.32}}
\def\piresultqsqsumcomb{  \ensuremath{1.13 \pm 0.14}}

\def\piresultqsqabcomb{  \ensuremath{0.82 \pm 0.12}}

\def\pipsyserrqsqafit{  \ensuremath{0.02}}
\def\pizsyserrqsqafit{  \ensuremath{0.01}}
\def\rhopsyserrqsqafit{ \ensuremath{0.04}}
\def\rhozsyserrqsqafit{ \ensuremath{0.03}}
\def\omegasyserrqsqafit{\ensuremath{0.01}}

\def\pipsyserrqsqbfit{  \ensuremath{0.02}}
\def\pizsyserrqsqbfit{  \ensuremath{0.01}}
\def\rhopsyserrqsqbfit{ \ensuremath{0.05}}
\def\rhozsyserrqsqbfit{ \ensuremath{0.03}}
\def\omegasyserrqsqbfit{\ensuremath{0.03}}

\def\pipsyserrqsqcfit{  \ensuremath{0.01}}
\def\pizsyserrqsqcfit{  \ensuremath{0.01}}
\def\rhopsyserrqsqcfit{ \ensuremath{0.04}}
\def\rhozsyserrqsqcfit{ \ensuremath{0.02}}
\def\omegasyserrqsqcfit{\ensuremath{0.01}}
\def\pisyserrqsqccomb{  \ensuremath{0.02}}

\def\pipsyserrqsqsumfit{  \ensuremath{0.05}}
\def\pizsyserrqsqsumfit{  \ensuremath{0.03}}
\def\rhopsyserrqsqsumfit{ \ensuremath{0.12}}
\def\rhozsyserrqsqsumfit{ \ensuremath{0.07}}
\def\omegasyserrqsqsumfit{\ensuremath{0.05}}
\def\pisyserrqsqsumcomb{  \ensuremath{0.06}}

\def\pisyserrqsqabcomb{  \ensuremath{0.04}}

\def\pipshortresultqsqafit{  \ensuremath{\pipresultqsqafit\ \pm
                                         \pipsyserrqsqafit}}
\def\pizshortresultqsqafit{  \ensuremath{\pizresultqsqafit\ \pm
                                         \pizsyserrqsqafit}}
\def\rhopshortresultqsqafit{ \ensuremath{\rhopresultqsqafit\ \pm
                                         \rhopsyserrqsqafit}}
\def\rhozshortresultqsqafit{ \ensuremath{\rhozresultqsqafit\ \pm
                                         \rhozsyserrqsqafit}}
\def\omegashortresultqsqafit{\ensuremath{\omegaresultqsqafit\ \pm
                                         \omegasyserrqsqafit}}

\def\pipshortresultqsqbfit{  \ensuremath{\pipresultqsqbfit\ \pm
                                         \pipsyserrqsqbfit}}
\def\pizshortresultqsqbfit{  \ensuremath{\pizresultqsqbfit\ \pm
                                         \pizsyserrqsqbfit}}
\def\rhopshortresultqsqbfit{ \ensuremath{\rhopresultqsqbfit\ \pm
                                         \rhopsyserrqsqbfit}}
\def\rhozshortresultqsqbfit{ \ensuremath{\rhozresultqsqbfit\ \pm
                                         \rhozsyserrqsqbfit}}
\def\omegashortresultqsqbfit{\ensuremath{\omegaresultqsqbfit\ \pm
                                         \omegasyserrqsqbfit}}

\def\pipshortresultqsqcfit{  \ensuremath{\pipresultqsqcfit\ \pm
                                         \pipsyserrqsqcfit}}
\def\pizshortresultqsqcfit{  \ensuremath{\pizresultqsqcfit\ \pm
                                         \pizsyserrqsqcfit}}
\def\rhopshortresultqsqcfit{ \ensuremath{\rhopresultqsqcfit\ \pm
                                         \rhopsyserrqsqcfit}}
\def\rhozshortresultqsqcfit{ \ensuremath{\rhozresultqsqcfit\ \pm
                                         \rhozsyserrqsqcfit}}
\def\omegashortresultqsqcfit{\ensuremath{\omegaresultqsqcfit\ \pm
                                         \omegasyserrqsqcfit}}

\def\pipshortresultqsqsumfit{  \ensuremath{\pipresultqsqsumfit\ \pm
                                           \pipsyserrqsqsumfit}}
\def\pizshortresultqsqsumfit{  \ensuremath{\pizresultqsqsumfit\ \pm
                                           \pizsyserrqsqsumfit}}
\def\rhopshortresultqsqsumfit{ \ensuremath{\rhopresultqsqsumfit\ \pm
                                           \rhopsyserrqsqsumfit}}
\def\rhozshortresultqsqsumfit{ \ensuremath{\rhozresultqsqsumfit\ \pm
                                           \rhozsyserrqsqsumfit}}
\def\omegashortresultqsqsumfit{\ensuremath{\omegaresultqsqsumfit\ \pm
                                           \omegasyserrqsqsumfit}}

\def\DpResultQsqSumFit{  \ensuremath{2.05 \pm 0.18}}
\def\DzResultQsqSumFit{  \ensuremath{2.07 \pm 0.17}}
\def\DspResultQsqSumFit{ \ensuremath{4.91 \pm 0.48}}
\def\DszResultQsqSumFit{ \ensuremath{5.16 \pm 0.41}}

\def\DpResultPDG{  \ensuremath{2.08 \pm 0.18}}
\def\DzResultPDG{  \ensuremath{2.15 \pm 0.22}}
\def\DspResultPDG{ \ensuremath{5.29 \pm 0.19}}
\def\DszResultPDG{ \ensuremath{6.5 \pm 0.5}}

\begin{document}


\preprint{\vbox{ \hbox{   }
                 \hbox{BELLE-CONF-0843}
                 \hbox{December 8, 2008}
}}

\title{ \quad\\[0.5cm]  Measurement of exclusive $B\to X_u \ell \nu$
  decays using full-reconstruction tagging at Belle}


\affiliation{Budker Institute of Nuclear Physics, Novosibirsk}
\affiliation{Chiba University, Chiba}
\affiliation{University of Cincinnati, Cincinnati, Ohio 45221}
\affiliation{Department of Physics, Fu Jen Catholic University, Taipei}
\affiliation{Justus-Liebig-Universit\"at Gie\ss{}en, Gie\ss{}en}
\affiliation{The Graduate University for Advanced Studies, Hayama}
\affiliation{Gyeongsang National University, Chinju}
\affiliation{Hanyang University, Seoul}
\affiliation{University of Hawaii, Honolulu, Hawaii 96822}
\affiliation{High Energy Accelerator Research Organization (KEK), Tsukuba}
\affiliation{Hiroshima Institute of Technology, Hiroshima}
\affiliation{University of Illinois at Urbana-Champaign, Urbana, Illinois 61801}
\affiliation{Institute of High Energy Physics, Chinese Academy of Sciences, Beijing}
\affiliation{Institute of High Energy Physics, Vienna}
\affiliation{Institute of High Energy Physics, Protvino}
\affiliation{Institute for Theoretical and Experimental Physics, Moscow}
\affiliation{J. Stefan Institute, Ljubljana}
\affiliation{Kanagawa University, Yokohama}
\affiliation{Korea University, Seoul}
\affiliation{Kyoto University, Kyoto}
\affiliation{Kyungpook National University, Taegu}
\affiliation{\'Ecole Polytechnique F\'ed\'erale de Lausanne (EPFL), Lausanne}
\affiliation{Faculty of Mathematics and Physics, University of Ljubljana, Ljubljana}
\affiliation{University of Maribor, Maribor}
\affiliation{University of Melbourne, School of Physics, Victoria 3010}
\affiliation{Nagoya University, Nagoya}
\affiliation{Nara Women's University, Nara}
\affiliation{National Central University, Chung-li}
\affiliation{National United University, Miao Li}
\affiliation{Department of Physics, National Taiwan University, Taipei}
\affiliation{H. Niewodniczanski Institute of Nuclear Physics, Krakow}
\affiliation{Nippon Dental University, Niigata}
\affiliation{Niigata University, Niigata}
\affiliation{University of Nova Gorica, Nova Gorica}
\affiliation{Osaka City University, Osaka}
\affiliation{Osaka University, Osaka}
\affiliation{Panjab University, Chandigarh}
\affiliation{Peking University, Beijing}
\affiliation{Princeton University, Princeton, New Jersey 08544}
\affiliation{RIKEN BNL Research Center, Upton, New York 11973}
\affiliation{Saga University, Saga}
\affiliation{University of Science and Technology of China, Hefei}
\affiliation{Seoul National University, Seoul}
\affiliation{Shinshu University, Nagano}
\affiliation{Sungkyunkwan University, Suwon}
\affiliation{University of Sydney, Sydney, New South Wales}
\affiliation{Tata Institute of Fundamental Research, Mumbai}
\affiliation{Toho University, Funabashi}
\affiliation{Tohoku Gakuin University, Tagajo}
\affiliation{Tohoku University, Sendai}
\affiliation{Department of Physics, University of Tokyo, Tokyo}
\affiliation{Tokyo Institute of Technology, Tokyo}
\affiliation{Tokyo Metropolitan University, Tokyo}
\affiliation{Tokyo University of Agriculture and Technology, Tokyo}
\affiliation{Toyama National College of Maritime Technology, Toyama}
\affiliation{Virginia Polytechnic Institute and State University, Blacksburg, Virginia 24061}
\affiliation{Yonsei University, Seoul}
  \author{I.~Adachi}\affiliation{High Energy Accelerator Research Organization (KEK), Tsukuba} 
  \author{H.~Aihara}\affiliation{Department of Physics, University of Tokyo, Tokyo} 
  \author{D.~Anipko}\affiliation{Budker Institute of Nuclear Physics, Novosibirsk} 
  \author{K.~Arinstein}\affiliation{Budker Institute of Nuclear Physics, Novosibirsk} 
  \author{T.~Aso}\affiliation{Toyama National College of Maritime Technology, Toyama} 
  \author{V.~Aulchenko}\affiliation{Budker Institute of Nuclear Physics, Novosibirsk} 
  \author{T.~Aushev}\affiliation{\'Ecole Polytechnique F\'ed\'erale de Lausanne (EPFL), Lausanne}\affiliation{Institute for Theoretical and Experimental Physics, Moscow} 
  \author{T.~Aziz}\affiliation{Tata Institute of Fundamental Research, Mumbai} 
  \author{S.~Bahinipati}\affiliation{University of Cincinnati, Cincinnati, Ohio 45221} 
  \author{A.~M.~Bakich}\affiliation{University of Sydney, Sydney, New South Wales} 
% ...
\collaboration{The Belle Collaboration}
\noaffiliation

\begin{abstract}
We report on a study of the branching fractions for the exclusive charmless
semileptonic $B$ decay modes \bpiplnu, \bpizlnu, \brhoplnu, \brhozlnu\ and
\bomegalnu,
using events tagged by fully reconstructing one of the $B$ mesons in a 
hadronic decay mode.  The obtained branching fractions are
$\mathcal{B}\left( \bpiplnu \right)
= \left( \pipshortresultqsqsumfit \right) \times 10^{-4}$,
$\mathcal{B}\left( \bpizlnu \right)
= \left( \pizshortresultqsqsumfit \right) \times 10^{-4}$,
$\mathcal{B}\left( \brhoplnu \right)
= \left( \rhopshortresultqsqsumfit \right) \times 10^{-4}$, 
$\mathcal{B}\left( \brhozlnu \right)
= \left( \rhozshortresultqsqsumfit \right) \times 10^{-4}$ and
$\mathcal{B}\left( \bomegalnu \right)
= \left( \omegashortresultqsqsumfit \right) \times 10^{-4}$,
where the first error in each case is statistical and the second 
systematic. Combining the charged and neutral pion modes using isospin 
invariance, the branching fraction obtained is
$\mathcal{B}\left( \bpilnu \right)
= \left( \piresultqsqsumcomb \pm \pisyserrqsqsumcomb \right)
\times 10^{-4}$. 
The partial branching fractions as a function of $q^2$ are extracted using 
three $q^2$ bins. At low $q^2$, the combined charged and neutral pion branching fractions and a Light Cone Sum Rules prescription imply
$|V_{ub}| = \left( 3.1 \pm 0.2 \pm 0.1 ^{+0.5}_{-0.3} \right) \times 10^{-3}$,
while using the high $q^2$ data and two different lattice prescriptions implies
$|V_{ub}| = \left( 3.1 \pm 0.3 \pm 0.1 ^{+0.6}_{-0.4} \right) \times 10^{-3}$ (HPQCD) and
$|V_{ub}| = \left( 3.3 \pm 0.4 \pm 0.1 ^{+0.6}_{-0.4} \right) \times 10^{-3}$ (FNAL) respectively.
In each case the errors are statistical, systematic and theoretical (associated
with the prescription used).  These results are obtained from a data sample 
that contains 657 $\times 10^6\ B\bar{B}$ pairs, collected near the 
$\Upsilon(4S)$ resonance with the Belle detector at the KEKB asymmetric energy 
$e^+ e^-$ collider. All results are preliminary.
\end{abstract}


\maketitle

\tighten

{\renewcommand{\thefootnote}{\fnsymbol{footnote}}}
\setcounter{footnote}{0}


\section{Introduction}

The Standard Model (SM) of particle physics contains a number of
parameters whose values are not predicted by theory and must therefore
be measured by experiment. In the quark sector, the elements of the
Cabibbo-Kobayashi-Maskawa (CKM) matrix \cite{CKM} govern
the weak transitions between quark flavours, and precision
measurements of their values are desirable. In particular, much
experimental and theoretical effort is currently being employed to
test the consistency of the Unitarity Triangle relevant to the decays
of $B$ mesons, which arises from the CKM formalism.

The angle $\sin 2\phi_1$, characterising indirect $CP$ violation in
$b \to c \overline{c} s$ transitions, is now known to a precision of
less than 4\% \cite{vervink-fpcp08}. This makes a precision
measurement of the length of the side of the Unitarity triangle
opposite to $\sin 2\phi_1$ particularly important as a consistency
check of the SM picture. The length of this side is determined to good
approximation by the ratio of the magnitudes of two CKM matrix
elements, $|V_{ub}|/|V_{cb}|$. Both of these can be measured using
exclusive semileptonic $B$ meson decays. Using charmed semileptonic
decays, the precision to which $|V_{cb}|$ has been determined is of
order 1.5\%. On the other hand $|V_{ub}|$,
which can be measured using charmless semileptonic decays, is 
poorly known by comparison. Both inclusive and exclusive
methods of measuring $|V_{ub}|$ have been pursued, with the inclusive
methods giving values approaching 5\% precision. The exclusive
determination of $|V_{ub}|$ currently has a precision closer to
10\%. It is the aim of an ongoing programme of measurements at the
$B$ factories to improve this precision to better than 5\%, for
comparison with the inclusive results, which have somewhat different
experimental and theoretical systematics, and to provide a sharp
consistency test with the value of $\sin 2\phi_1$.

Measurements of branching fractions for exclusive \bxulnu\ decays,
where $X_u$ denotes a light meson containing a $u$ quark, have been
reported by the
  CLEO~\cite{cleo-untagged-00,
             cleo-untagged-03,
             cleo-untagged-07},
 BaBar~\cite{babar-untagged-03,
             babar-untagged-05,
             babar-untagged-07,
             babar-sl-08,
             babar-fullrecon-06,
             babar-fullrecon-eta-06}
and
 Belle~\cite{belle-untagged-omega-04,
             belle-sl-07,
             belle-fullrecon-06}
collaborations. A recent compilation of these results has been made by
the Heavy Flavour Averaging Group (HFAG) \cite{HFAG}. In these
measurements, three methods of identifying signal candidates have been
employed, denoted ``untagged'', ``semileptonic tagged'' or
``full reconstruction tagged''.

The most precise measurements at the present time come from
untagged analyses \cite{cleo-untagged-07} \cite{babar-untagged-07}.
As more integrated luminosity is accumulated by the $B$-factory
experiments, full reconstruction tagging techniques to identify
candidate $B$ mesons, against which the signal $B$ mesons recoil,
will become the most precise method. These techniques hold the advantage
of providing the best signal to background ratio, offset by the lowest
efficiencies. In this paper we present preliminary studies of the exclusive
semileptonic decays \bpiplnu, \bpizlnu, \brhoplnu, \brhozlnu\ and
\bomegalnu\ using such a full reconstruction tagging technique .


\section{Experimental Procedure}

\subsection{Data Sample and the Belle Detector}

The data sample used for this analysis contains
657 $\times 10^6 B\overline{B}$ pairs, collected  with the
Belle detector at the KEKB asymmetric-energy $e^+e^-$ (3.5 on 8~GeV)
collider~\cite{KEKB}. KEKB operates at the $\Upsilon(4S)$ resonance 
($\sqrt{s}=10.58$~GeV) with a peak luminosity that exceeds
$1.7\times 10^{34}~{\rm cm}^{-2}{\rm s}^{-1}$. The $\Upsilon(4S)$ is
produced with a Lorentz boost of $\beta\gamma=0.425$ nearly along
the electron beamline ($z$).

The Belle detector is a large-solid-angle magnetic spectrometer that
consists of a silicon vertex detector (SVD), a 50-layer central drift
chamber (CDC), an array of aerogel threshold Cherenkov counters
(ACC), a barrel-like arrangement of time-of-flight scintillation
counters (TOF), and an electromagnetic calorimeter (ECL) comprised of
CsI(Tl) crystals located inside a super-conducting solenoid coil
that provides a 1.5~T magnetic field.  An iron flux-return located outside
of the coil is instrumented to detect $K_L^0$ mesons and to identify
muons (KLM).  The detector is described in detail elsewhere~\cite{Belle}.
Two inner detector configurations were used. A 2.0 cm beampipe
and a 3-layer silicon vertex detector was used for the first sample
of 152 $\times 10^6\ B\bar{B}$ pairs, while a 1.5 cm beampipe, a 4-layer
silicon detector and a small-cell inner drift chamber were used to record  
the remaining 505 $\times 10^6 B\bar{B}$ pairs\cite{svd}.


\subsection{Full Reconstruction Tagging}

In this analysis we fully reconstruct one of the two $B$ mesons from
the $\Upsilon(4S)$ decay (\btag) in one of the following hadronic
decay modes,
$B^- \to D^{(*)0} \pi^-$, $B^- \to D^{(*)0} \rho^-$, $B^- \to D^{(*)0} a_1^-$,
$B^- \to D^{(*)0} D_s^{(*)-}$,
$\bzerobar \to D^{(*)+} \pi^-$,
$\bzerobar \to D^{(*)+} \rho^-$,
$\bzerobar \to D^{(*)+} a_1^-$
or
$\bzerobar \to D^{(*)+} D_s^{(*)-}$.
Within these $B$ decay modes, the $D$ mesons used in the
reconstruction of \btag\ are
    $D^0 \to K^- \pi^+$, $K^- \pi^+ \pi^0$, $K^- \pi^+ \pi^- \pi^+$,
    $K_S^0 \pi^0$, $K_S^0 \pi^+ \pi^-$, $K_S^0 \pi^+ \pi^- \pi^0$ and
    $K^+ K^-$, $D^+ \to K^- \pi^+ \pi^+$, $K^- \pi^+ \pi^+ \pi^0$,
    $K_S^0 \pi^+$, $K_S^0 \pi^+ \pi^0$, $K_S^0 \pi^+ \pi^+ \pi^-$ and
    $K^+ K^- \pi^+$, and $D_s^- \to K_S^0 K^-$ and $K^+ K^- \pi^-$.
$D^{*}$ mesons are reconstructed by combining a $D$ candidate and a
soft pion or photon \cite{CC}.

The selection of \btag\ candidates is based on the proximity of the
beam-energy constrained mass
$\mbc = \sqrt{\ebeam^2 - \pb^2}$ and energy difference
$\deltae = E_B - \ebeam$ to their nominal values of the $B$ meson rest
mass and zero, respectively. Here \ebeam, \pb\ and $E_B$ are the beam
energy and the measured momentum and energy of the \btag\ candidate
in the $\Upsilon(4S)$ rest frame respectively. To be considered as a
candidate, loose preselection conditions of $5.2 < \mbc < 5.3\ \GeV\ $
and $-0.3 < \Delta E < 0.3\ \GeV\ $ must be satisfied. If an event has
multiple \btag\ candidates, these are ordered according to a $\chi^2$
variable based on $\Delta E$, the $D$ mass and the $D^*$ - $D$ mass
difference if appropriate. Following selection of the most likely
\btag\ candidate, events with \btag\ satisfying the tighter selection
criteria $\mbc > 5.27\ \GeV\ $ and $-0.08 < \deltae < 0.06\ \GeV\ $
are retained.


\subsection{Signal Reconstruction}

Reconstructed charged tracks and ECL clusters which are not associated with the
\btag\ candidate are used to search for the signal $B$ meson decays of
interest recoiling against the \btag. Photons identified with isolated
ECL clusters which have a laboratory energy of less than 50 MeV are ignored.
Electrons are identified using information on $dE/dx$ from the CDC,
response of the ACC, position matching between the reconstructed track
and an ECL cluster, the ECL shower shape and the ratio of the
energy deposited in the ECL to the momentum determined from tracking.
The signals in the KLM are used to identify muons. Charged kaons are
identified based on the $dE/dx$ information from the CDC, the
Cherenkov light yields in the ACC and time-of-flight information from
the TOF counters. Any charged particles which are not identified as
leptons or kaons are taken to be pions.
Photons whose direction in the laboratory frame lies within a $5^\circ$ cone 
of the direction of an identified electron or positron are considered to be
bremsstrahlung. The 4-momentum of the photon is added to that of the
lepton and the photon is not considered further.

Neutral pions are reconstructed from pairs of photons whose invariant
mass lies in the range $[0.120, 0.150]$\ \GeV, of order $\pm 3 \sigma$
of the $\pi^0$ mass. Charged  $\rho$ meson
candidates are reconstructed via the decay $\rho^\pm \to \pi^\pm$
$\pi^0$ where the invariant mass of the pair of pions is required to
lie in the range $[0.570, 0.970]$\ \GeV. Neutral $\rho$ meson candidates
are similarly reconstructed from pairs of oppositely charged pions,
with the requirement that $m_{\pi^+ \pi^-}$ is in the range
$[0.690, 0.850]$\ \GeV. Finally, $\omega$ candidates are
reconstructed from $\omega \to \pi^+$ $\pi^-$ $\pi^0$ with
$m_{\pi^+ \pi^- \pi^0}$ in the range $[0.703, 0.863]$\ \GeV.
In events where more than one hadron candidate, denoted $X_u$, of a
given type is identified amongst the recoil particles, the candidate
with the highest momentum in the $\Upsilon(4S)$ rest frame is chosen.

To isolate signal candidates, several requirements are placed on
the recoil system. There must be one lepton candidate present only.
The total charge of the recoil system, $Q_\mathrm{recoil}$, is required to
be $0$ if a neutral tag has been identified, and $\pm 1$ if a charged
tag has been identified. In the charged case, the sign of
$Q_\mathrm{recoil}$ must be opposite to that of $Q_\mathrm{tag}$. In
the neutral case, we do not make any requirement on the sign of the
lepton charge with respect to the \btag, to allow for mixing.

The number of charged recoil particles is required to correspond to
one of the sought signal modes, i.e. one for \bpizlnu\ (the lepton),
two for \bpiplnu\ and \brhoplnu\ (the lepton
plus one charged pion) and three for \brhozlnu\ and \bomegalnu\ (the
lepton plus two charged pions). Additionally, the number of recoil
$\pi^0$ candidates is required to be consistent with one of the sought
modes. In order to increase efficiency, however, we allow more than
the necessary number in some cases:  we require
no $\pi^0$ candidates to be present for \bpiplnu\ and \brhozlnu\ modes,
and at least one $\pi^0$ for the \bpizlnu\ and \brhoplnu\ modes, and
exactly one $\pi^0$ for the \bomegalnu\ mode. Additionally, we require
that there be no more than $0.5\ \GeV\ $ of residual neutral energy
present on the recoil side, calculated in the $\Upsilon(4S)$ rest
frame, after any photons contributing to the $X_u$ candidate have been
removed.

If the tagging $B$ is correctly reconstructed and the correct lepton
and hadron candidate have been identified on the recoil side, then
ideally all missing 4-momentum is due to the remaining unidentified neutrino.
Signal events can therefore identified by examining the missing mass squared
(\mm2), defined to be the square of the missing 4-momentum.
In signal events this quantity should be close to zero, and
applying this requirement provides a very strong discrimination
between signal and background. In practice we construct the recoiling
$B$ meson to have its nominal energy and magnitude of momentum in the
$\Upsilon(4S)$ rest frame, and direction opposite to \btag.
The missing 4-momentum vector of the decaying recoil $B$ system is then
obtained by subtracting the 4-momenta of the hadron candidate and
lepton from the $B$ 4-momentum.

For signal candidates, the neutrino 4-momentum is defined to be
$p_{\nu} = \left( |\vec{p}_\mathrm{miss}|, \vec{p}_\mathrm{miss} \right)$,
where $\vec{p}_\mathrm{miss}$ is the missing 3-momentum vector of the
recoil $B$ system defined as above. The kinematical variable
\qsq, defined to be the invariant mass squared of the lepton-neutrino
system, can then be determined. The \qsq\ resolution obtained using
this procedure is excellent, varying from $0.21\ \GeVGeV\ $ for the
\bpiplnu\ channel to $0.28\ \GeVGeV\ $ for the \bomegalnu\ channel.


\subsection{Background Estimation}

Background contributions come from several sources. These include
semileptonic decays resulting from \btoclnu\ transitions, denoted
\bxclnu, which have significantly larger branching fractions than the
channels under study; continuum $e^+ + e^- \to q\overline{q}$
processes; and cross feed from one \bxulnu\ channel into another.
The contributions of these backgrounds are studied
using Monte Carlo (MC) simulated data samples generated with the
EvtGen package \cite{evtgen}. Generic \bb and continuum MC samples
equivalent to approximately three times the integrated luminosity of
the real data set are used. The model
adopted for \bdstarlnu\ and \bdlnu\ decays is based on HQET and
parametrisation of the form factors \cite{caprini}, while \bddstarlnu\
decays are based on the ISGW2 model \cite{isgw2}. A non-resonant
$B \to D^{(*)}\pi \ell \nu$ component based on the Goity-Roberts
prescription \cite{goity} is also included.

A separate MC sample equivalent to approximately sixteen times the
integrated luminosity of the real data set is used to simulate the signal
channels and crossfeed from other \bxulnu\ decays. Models for the
exclusive modes are based on Light Cone Sum Rules (LCSR) for
$\pi$ \cite{ball-and-zwicky-01}, $\rho$ and $\omega$
\cite{ball-and-braun-98} modes and ISGW2 \cite{isgw2}
for other exclusive modes.

Radiative effects associated with the lepton and resulting from
higher-order QED processes are modelled in all MC samples using the
PHOTOS package \cite{photos}. All generated MC events are passed
through a full simulation of Belle detector effects based on GEANT
3.21 \cite{geant}.


\section{Results and Systematic Uncertainties}


\subsection{Signal yield determination}

In order to obtain the signal yields, a fit is performed to the
observed \mm2\ distributions, individually for each mode. The fits
are made in three separate bins of \qsq\ in order to obtain the yields
as a function of \qsq, and to minimise the systematic error which
arises from the lack of precise knowledge of the shape of the form
factors in \bxulnu\ decays. The \qsq\ bins are chosen commensurate
with available statistics, and are
$0 \mathrm{\ to \ } 8\ \GeVGeV$,
$8 \mathrm{\ to \ } 16\ \GeVGeV$, and greater than $16\ \GeVGeV$.
The components of any given fit are signal, \ulnu\ crossfeed and the
contribution from other backgrounds, which is dominated by \bxclnu\
decays. The shapes of the components are taken from MC and the
normalizations are fit parameters. The fitting method follows that of
Barlow and Beeston \cite{barlow-and-beeston-93} and takes into account
finite MC statistics. In Fig.~\ref{fig:mm2fit} the observed
\mm2\ distributions for the five decay modes are shown, summed over
the three \qsq\ bins. The fit components shown in the figures are
likewise those obtained by summing the results of the fits in the
individual \qsq\ bins. The fitted event yields obtained in this way
are $59 \pm 10$ for the \bpiplnu\ mode, $49 \pm 9$ for \bpizlnu,
$65 \pm 12$ for \brhoplnu, $80 \pm 10$ for \brhozlnu\ and $25 \pm 8$
for \bomegalnu.


\subsection{Branching Fractions}

\begin{figure}[p]
\caption{Missing mass squared (\mm2)
  distributions after all selection criteria,
  for (a) \bpiplnu, (b) \bpizlnu, (c) \brhoplnu, (d) \brhozlnu,
and (e) \bomegalnu\ modes.
Data is indicated by the points with error
bars. The signal histogram (lightest shade in greyscale in each case) shows the
fitted prediction based on the LCSR
model \cite{ball-and-zwicky-01, ball-and-braun-98}.
The green histogram (middle shade in greyscale) shows the fitted
$b \to u \ell \nu$ background
contribution. The crimson histogram (darkest shade in greyscale) shows
the fitted background contribution from other sources. The fitting
method is explained in the text.}
\label{fig:mm2fit}
\begin{tabular}{ll}
  (a) & (b) \\
\includegraphics[width=0.35\textwidth]{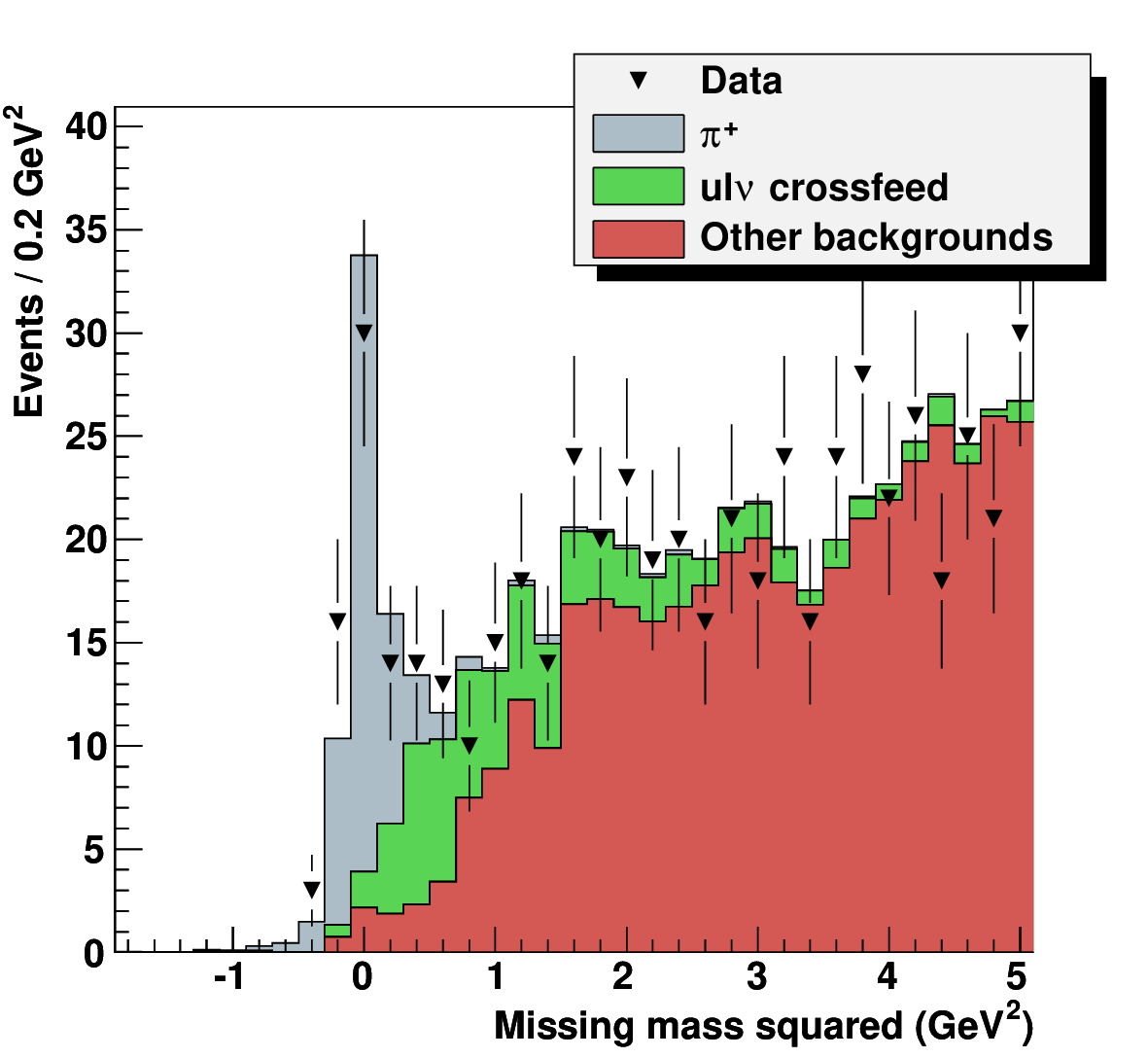} &
\includegraphics[width=0.35\textwidth]{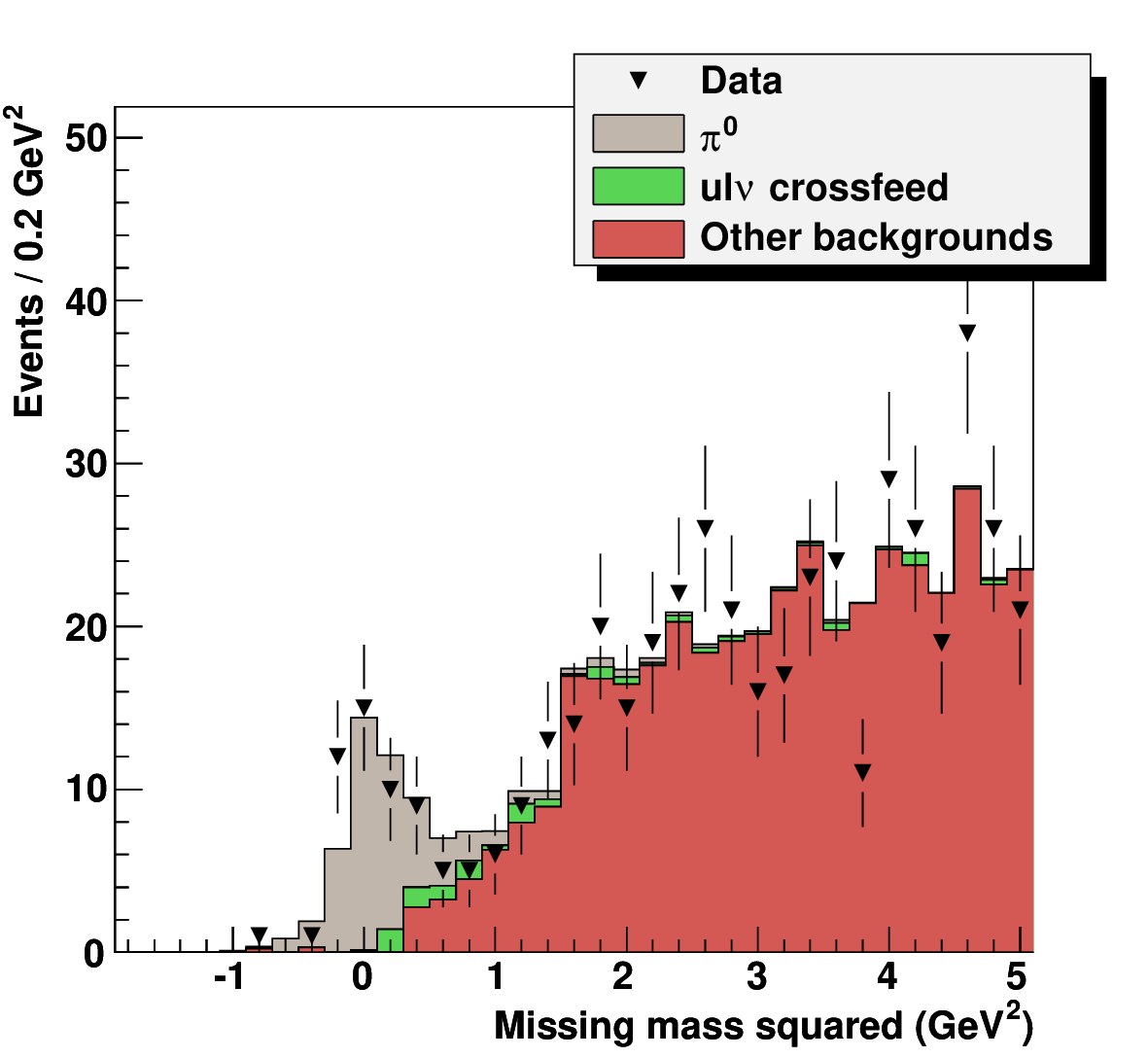} \\
  (c) & (d) \\
\includegraphics[width=0.35\textwidth]{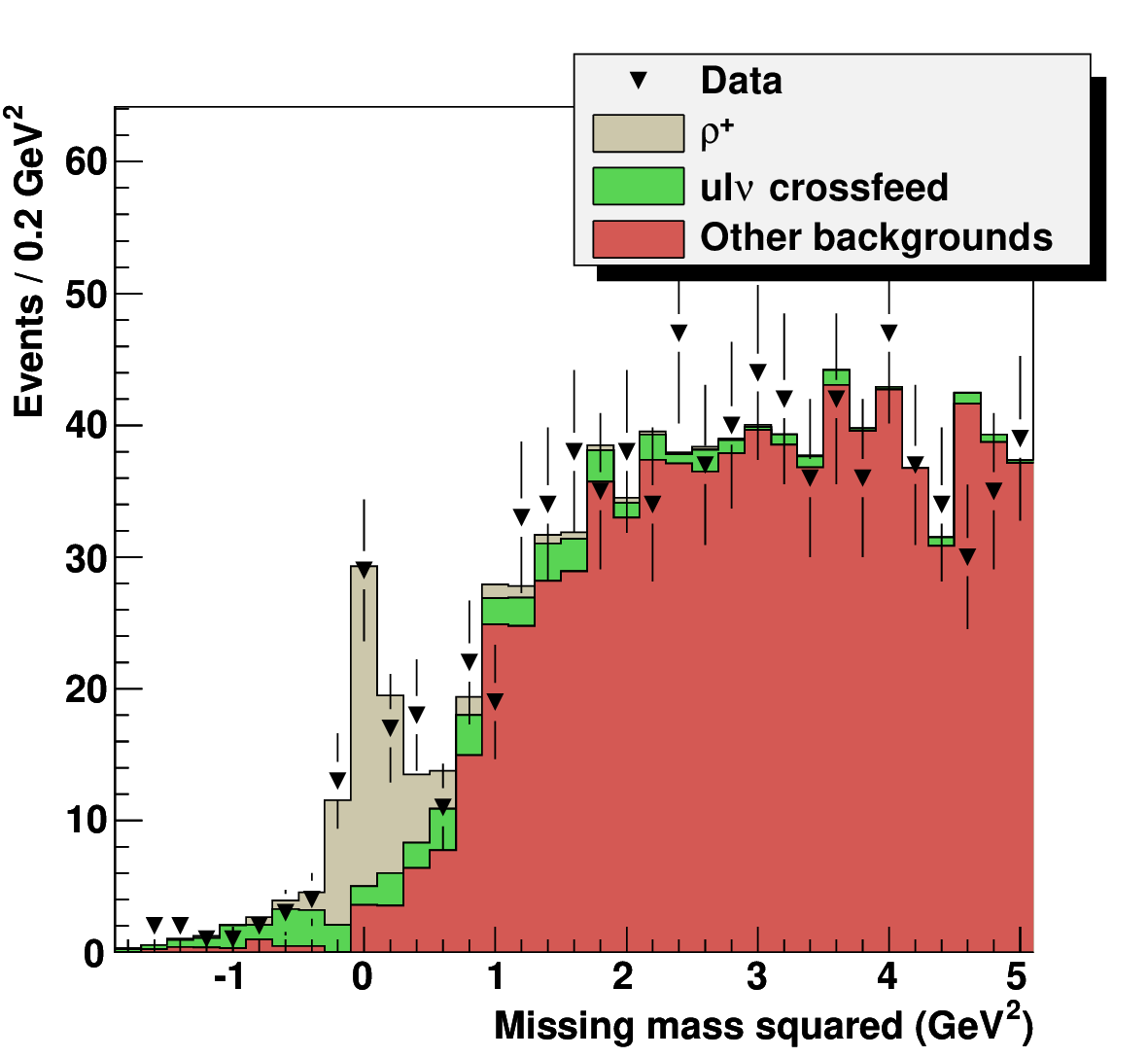} &
\includegraphics[width=0.35\textwidth]{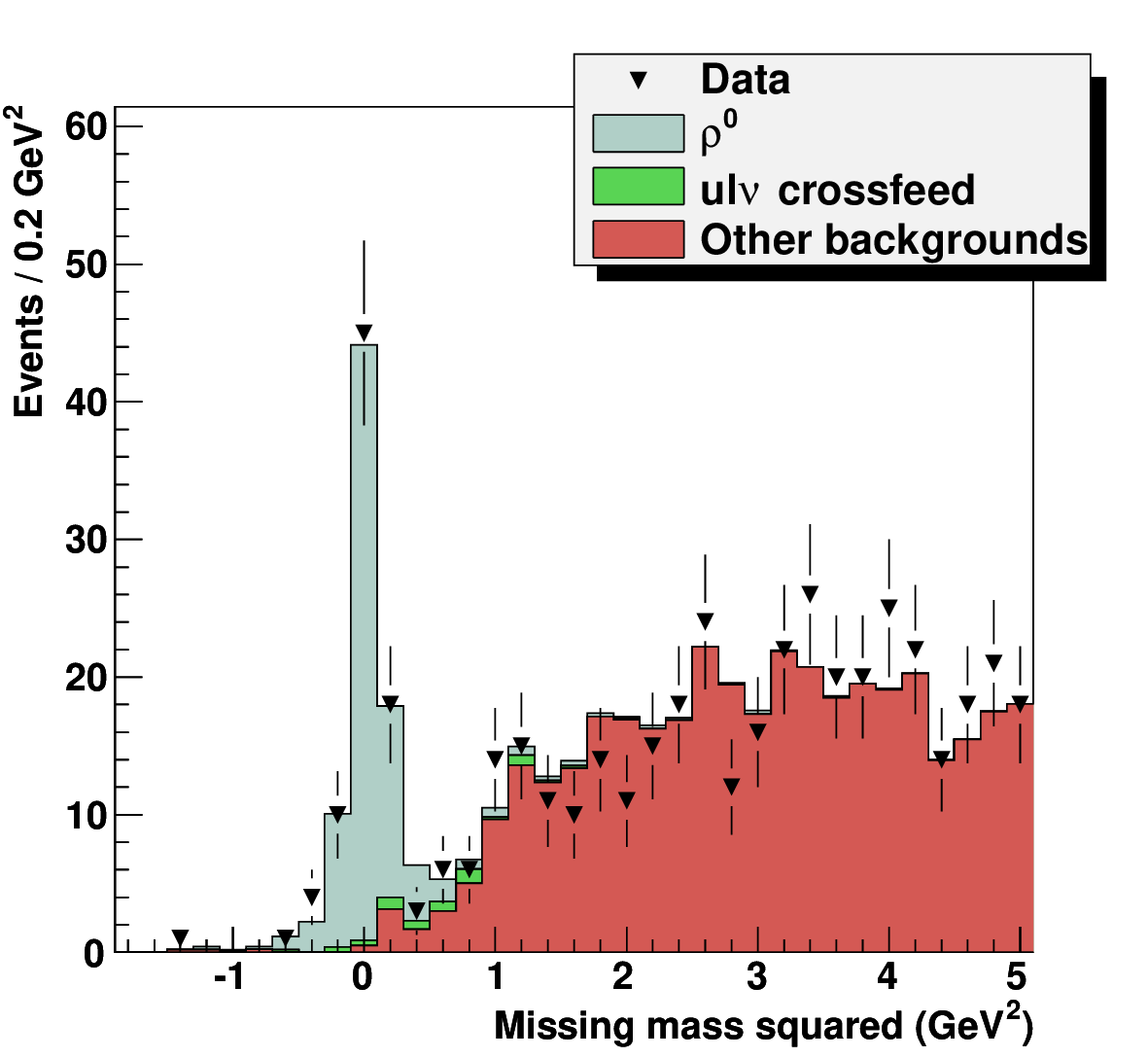} \\
  (e) &     \\
\includegraphics[width=0.35\textwidth]{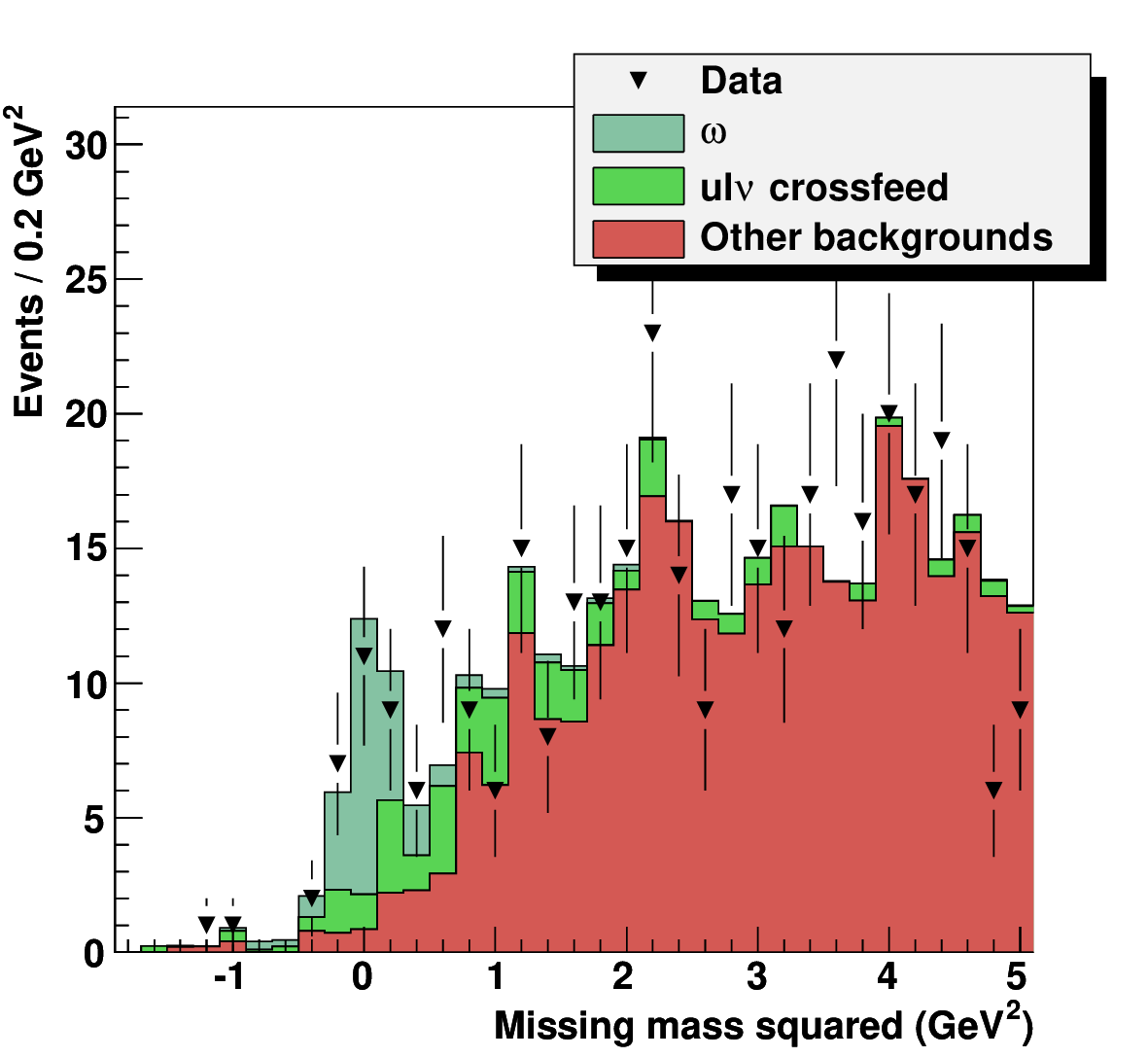} &
\end{tabular}
\end{figure}

We extract the partial branching fractions in bins of \qsq, using the
formula
\begin{eqnarray*}
  \Delta \mathcal{B}(\bxulnu) = \dfrac{\mathrm{Signal\ Yield}}
	                       {\epsilon \cdot 2 N_{\bb}}
\end{eqnarray*}
where $\epsilon$ is the signal efficiency within the \mm2\ range
corresponding to the histograms in Fig.~\ref{fig:mm2fit}, and
$N_{\bb}$ is the number of $B\bar{B}$ pairs which the data set is
estimated to contain before any event selections are made.
The signal efficiencies are estimated from MC and do
not exhibit a strong \qsq\ dependence. Averaged over \qsq, these
efficiencies are
(\pipeffglobal)\% for the \bpiplnu\ mode, (\pizeffglobal)\% for
\bpizlnu, (\rhopeffglobal)\% for \brhoplnu, (\rhozeffglobal)\% for
\brhozlnu\ and (\omegaeffglobal)\% for \bomegalnu. The number of
$B\bar{B}$ pairs is $(656.6 \pm 8.9) \times 10^6$.

\begin{figure}[p]
\caption{Partial branching fractions as a function of \qsq\ for the
  five signal modes (a) \bpiplnu, (b) \bpizlnu, (c) \brhoplnu,
  (d) \brhozlnu, and (e) \bomegalnu. Errors shown are statistical and
  preliminary systematic, added in quadrature. LCSR predictions
  \cite{ball-and-zwicky-01, ball-and-braun-98} are shown in blue
  (solid line) and a quark model prediction \cite{isgw2} in red
  (dashed line). The predictions are normalised to have the
  same area as that of the data distribution in each case.}
\label{fig:qsq_bfs}
  \begin{tabular}{ll}
    (a) & (b) \\
\includegraphics[width=0.35\textwidth]{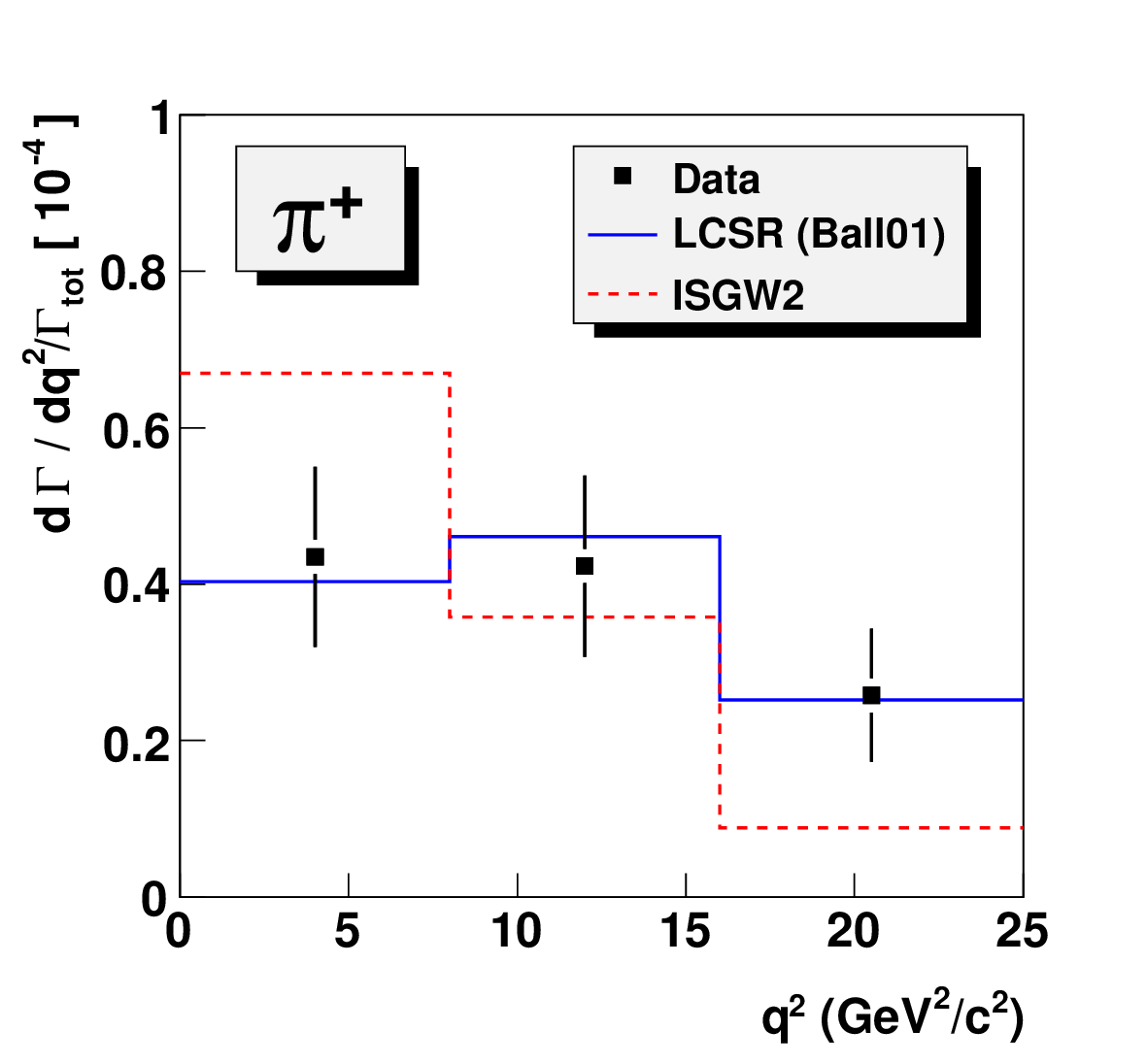}  &
\includegraphics[width=0.35\textwidth]{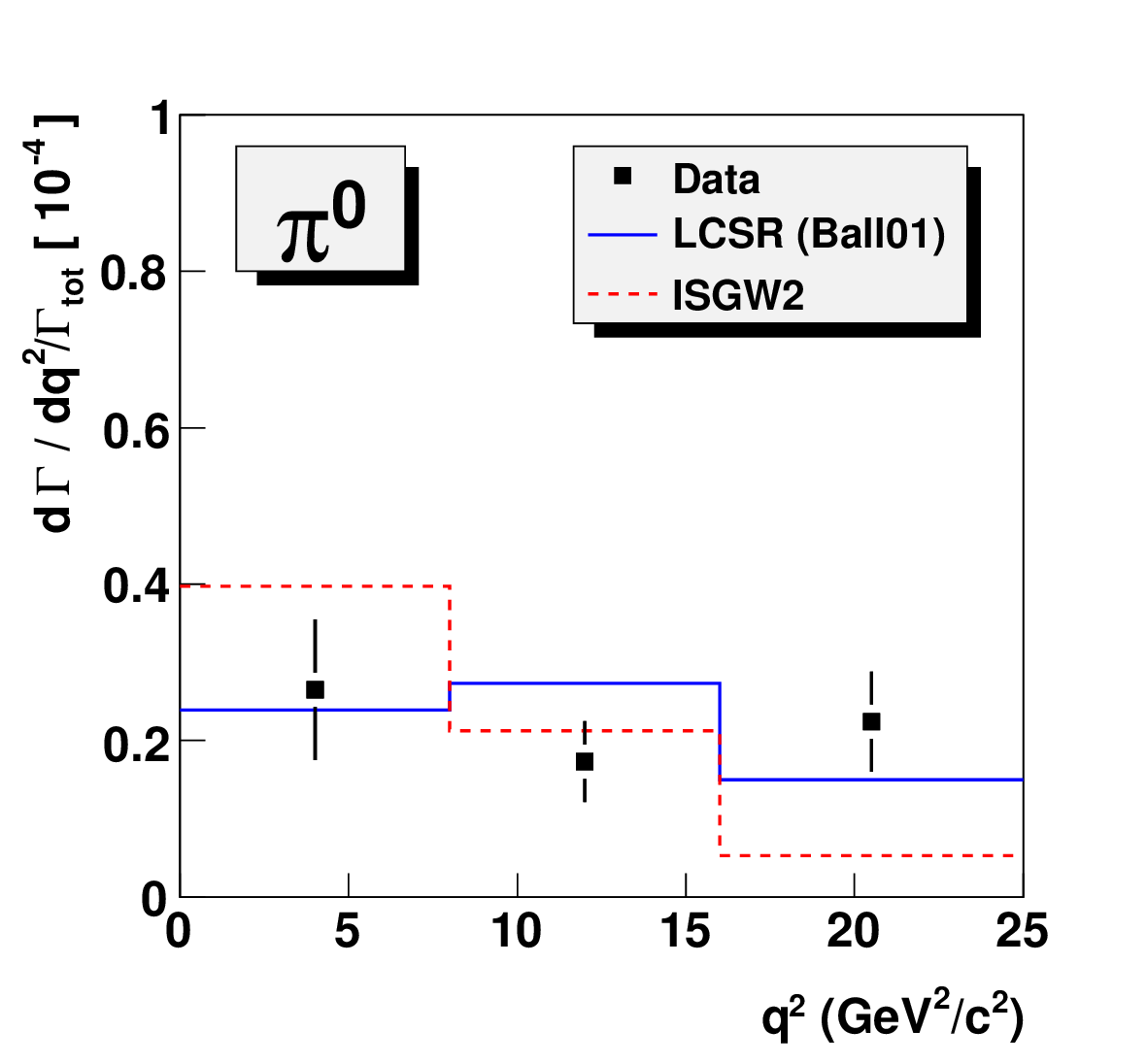}  \\
    (c) & (d) \\
\includegraphics[width=0.35\textwidth]{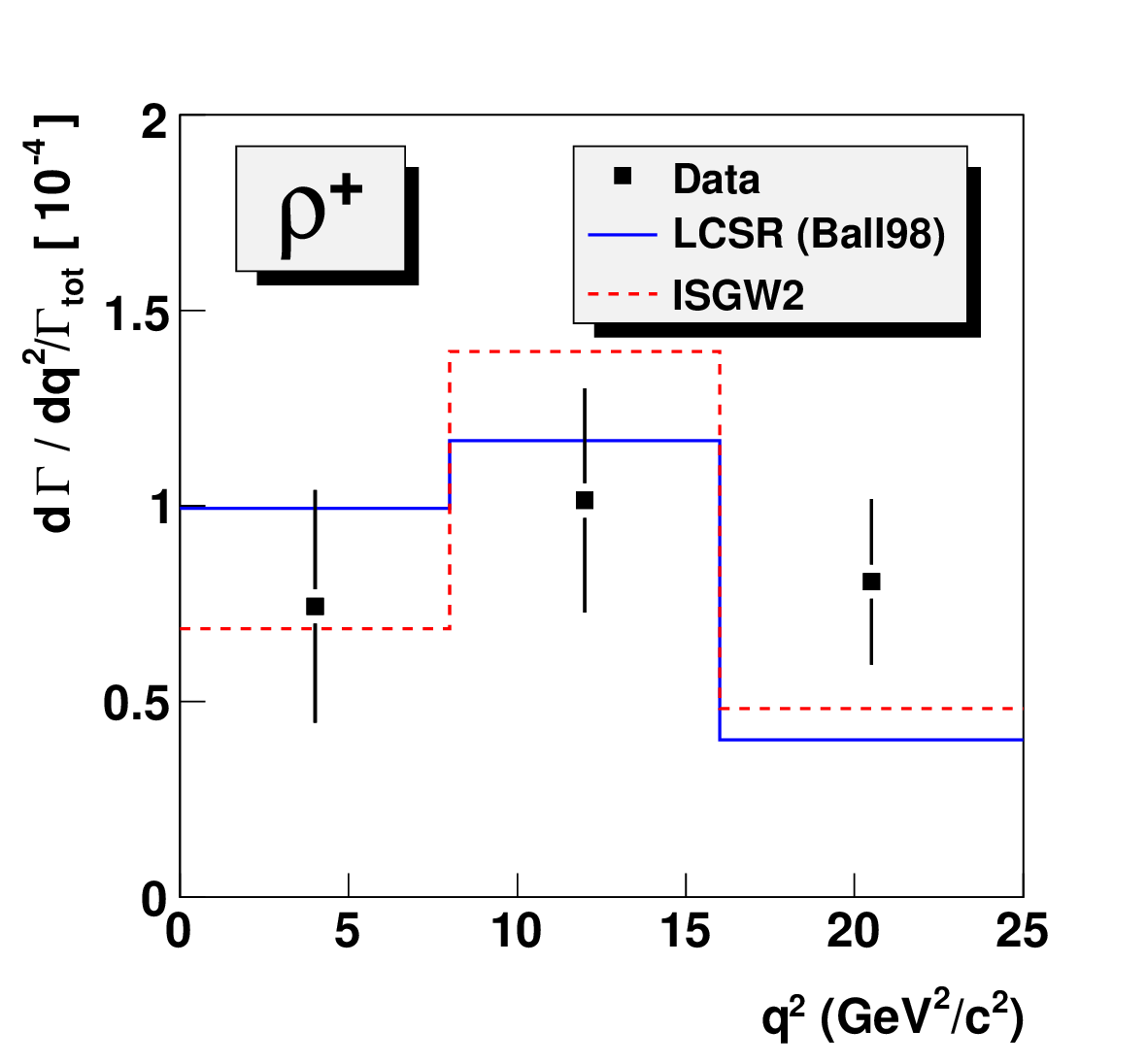}  &
\includegraphics[width=0.35\textwidth]{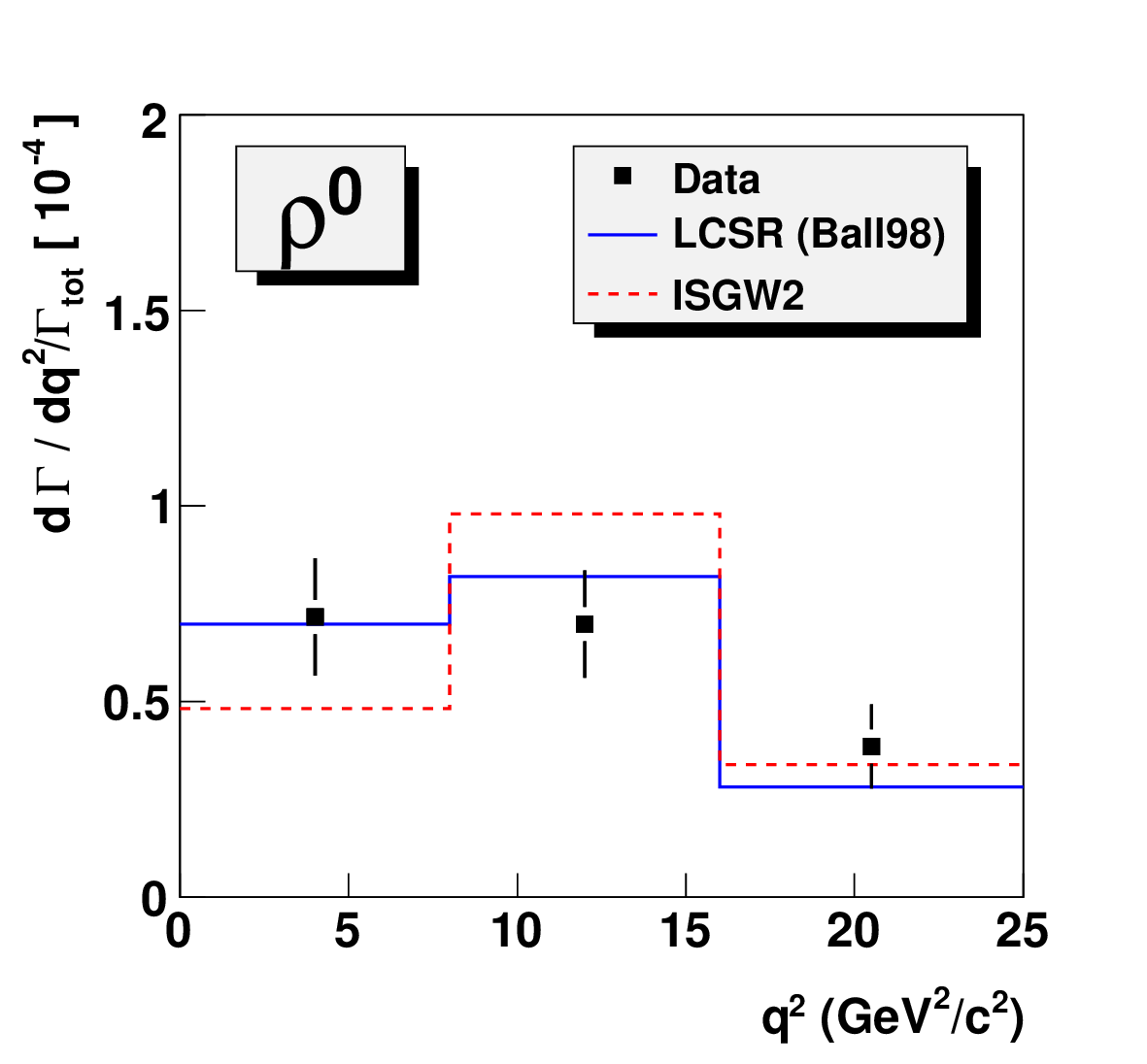}  \\
    (e) &     \\
\includegraphics[width=0.35\textwidth]{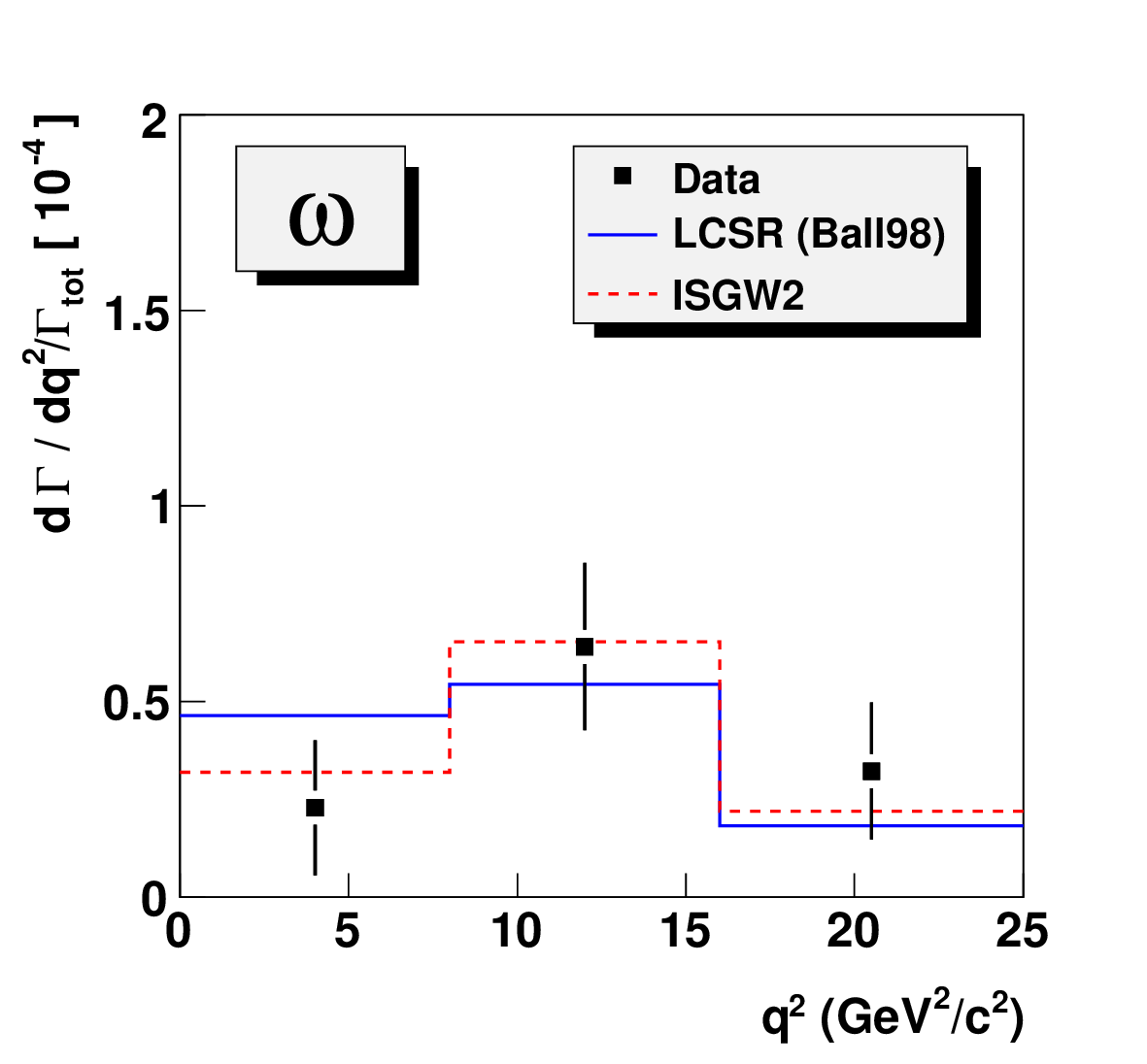}  \\
\end{tabular}
\end{figure}

The resultant partial branching fractions are given in
Table~\ref{table:qsq_bfs}. The first error given in each case is
statistical, and the second systematic, as described in the following
section. Figure~\ref{fig:qsq_bfs} presents the shapes of the
partial branching fractions for all five modes as a function of \qsq,
where the error bars displayed are obtained by adding the statistical
and systematic errors in quadrature. Also shown in this figure are the
predictions based on LCSR prescriptions
\cite{ball-and-zwicky-01, ball-and-braun-98} and a quark model
prescription \cite{isgw2}. The predictions are normalised to have the
same area as that of the data distribution in each case.
For the pion decay modes in particular it is clear that the LCSR
prescription is in better agreement with the data. 
\begin{table}[htb]
    \caption{Partial branching fractions in three
      bins of \qsq. These are summed to give the full branching
      fraction quoted in the ``Sum'' column. Errors are statistical
      and systematic.}
    \label{table:qsq_bfs}
    \begin{center}
    \begin{tabular}
      {@{\hspace{0.3cm}}c@{\hspace{0.3cm}}  
       @{\hspace{0.3cm}}c@{\hspace{0.3cm}}  
       @{\hspace{0.3cm}}c@{\hspace{0.3cm}}  
       @{\hspace{0.3cm}}c@{\hspace{0.3cm}}  
       @{\hspace{0.3cm}}c@{\hspace{0.3cm}}}
      \hline \hline
      &
      \multicolumn{3}{c}{$\Delta\mathcal{B} \left[ 10^{-4}\right]$} &
      {$\mathcal{B} \left[ 10^{-4}\right]$} \\
      \cline{2-5}
      Mode
      & $0 < q^2 < 8$
      & $8 < q^2 < 16$
      & $q^2 > 16$
      & Sum \\
      & (\GeVGeV)
      & (\GeVGeV)
      & (\GeVGeV)
      & (\GeVGeV) \\
      \hline 
      \bpiplnu
      & \pipshortresultqsqafit 
      & \pipshortresultqsqbfit
      & \pipshortresultqsqcfit
      & \pipshortresultqsqsumfit \\
      \bpizlnu
      & \pizshortresultqsqafit
      & \pizshortresultqsqbfit
      & \pizshortresultqsqcfit
      & \pizshortresultqsqsumfit \\ 
      \brhoplnu
      & \rhopshortresultqsqafit
      & \rhopshortresultqsqbfit
      & \rhopshortresultqsqcfit
      & \rhopshortresultqsqsumfit \\ 
      \brhozlnu
      & \rhozshortresultqsqafit
      & \rhozshortresultqsqbfit
      & \rhozshortresultqsqcfit
      & \rhozshortresultqsqsumfit \\ 
      \bomegalnu
      & \omegashortresultqsqafit
      & \omegashortresultqsqbfit
      & \omegashortresultqsqcfit
      & \omegashortresultqsqsumfit \\ 
      \hline
    \end{tabular}
    \end{center}
\end{table}

\subsection{Systematic Uncertainties}

Table~\ref{table:systsum}
summarises the preliminary result of a study of the contributions to
the total systematic error for the branching fractions summed over the
three \qsq\ bins, for each of the \bxulnu\ signal modes.
These are broken down into the following categories;
those arising from detector simulation, such as charged track
reconstruction efficiency, particle identification and neutral cluster
reconstruction; uncertainties in the luminosity; and effects of the
form factor models used and assumed branching fractions in the MC.

The detector simulation errors have been obtained following the
procedure described in a Belle study of similar final states in
reference~\cite{belle-sl-07}.
The effects of model dependence of the form factor shapes assumed in the
\bxulnu\ MC used for signal efficiency and crossfeed background estimates
have been studied by comparing the fitted yields obtained using the
default model implemented in the MC, which is LCSR
\cite{ball-and-zwicky-01, ball-and-braun-98}, and the ISGW2
model \cite{isgw2}. This is achieved by reweighting the MC events
on an event-by-event basis based on their generated values of \qsq\
and angular variables.  The variation between these two models in
predicting the shapes of the \qsq\ distributions for the pseudoscalar
and vector modes typifies the spread between available models for the
dynamics of these decays. 

The shapes of the background \mm2\ components used in the fits can be
affected by the assumed branching fractions of dominantly contributing
\btoulnu\ and \btoclnu\ decays in the MC samples used.
This was studied by varying in turn the \bpiplnu, \bpizlnu, \brhoplnu,
\brhozlnu, \bomegalnu, \bdplnu, \bdzlnu, \bdsplnu\ and \bdszlnu\
branching fractions by their measurement errors as quoted by the
Particle Data Group~\cite{PDG}. A reweighting technique is again used,
and fitted yields with and without reweighting are compared. The maximum
observed spread in the fitted branching fraction is assigned as
systematic error.

The effects of finite MC statistics are taken into account in the
fitting procedure \cite{barlow-and-beeston-93} and are reflected in the
errors on the obtained branching fractions. Since the available MC
samples are rather limited in statistics, variations of the
assumptions on form factor shapes and normalizations can be absorbed
by the present fits to a significant extent.

\begin{table}[htp]
\caption{Preliminary results of a study of sources of systematic uncertainty.}
\label{table:systsum}
\begin{center}
\begin{tabular}
      {@{\hspace{0.35cm}}l@{\hspace{0.35cm}}  
       @{\hspace{0.35cm}}c@{\hspace{0.35cm}}  
       @{\hspace{0.35cm}}c@{\hspace{0.35cm}}  
       @{\hspace{0.35cm}}c@{\hspace{0.35cm}}  
       @{\hspace{0.35cm}}c@{\hspace{0.35cm}}  
       @{\hspace{0.35cm}}c@{\hspace{0.35cm}}}  
  \hline \hline
 Source of error & \multicolumn{5}{c}{Assigned systematic error} \\
 \hline
  & \bpiplnu & \bpizlnu & \brhoplnu & \brhozlnu & \bomegalnu \\
 \hline
     {\bf Detector Simulation:} & & & & & \\
     Pion track finding eff. & 1.0\% & - & 1.0\% & 2.0\% & 2.0\%  \\
     $\pi^0$ reconstruction eff. & - & 2.0\% & 2.0\% & - & 2.0\%  \\
     Lepton track finding eff. & 1\% & 1\% & 1\% & 1\% & 1\% \\
     Lepton identification & 2.1\% & 2.1\% & 2.1\% & 2.1\% & 2.1\% \\
     Pion identification & 2.0\% & 2.0\% & 2.0\% & 2.0\% & 2.0\% \\
     Combined & 3.2\% & 3.7\% & 3.8\% & 3.7\% & 4.2\% \\
     \hline
     N(\bb) uncertainty & 1.36\% & 1.36\% & 1.36\% & 1.36\% & 1.36\%  \\
     \hline 
	 {\bf Form Factor Shapes:}  & & & & & \\
	 $\pi$ (LCSR $\to$ ISGW2) & 1.5\% & 0.0\% & 1.0\% &  0.0\% & 0.0\% \\
	 $\rho,\omega$ (LCSR $\to$ ISGW2)
	                          & 1.8\% & 0.5\% & 2.2\% & 0.1\% & 0.2\% \\
     \hline 
	 {\bf Branching Fractions:} & & & & & \\
	 \btoulnu/\btoclnu\ norm. & 0.3\% & 0.2\% & 1.1\% & 0.6\% & 0.5\% \\
\hline 
 {\bf Total systematic error} &  4.2\% & 3.9\% & 4.8\% & 4.0\% & 4.4\% \\
  \hline \hline
\end{tabular}
\end{center}
\end{table}


\subsection{Fully reconstructed charmed semileptonic decays}

In order to check the robustness of our method, we apply it to
exclusive \bxclnu\ modes, where the branching fractions are larger and
better known than in the \bxulnu\ case. We use the channels \bdplnu,
\bdzlnu, \bdsplnu\ and \bdszlnu, applying a similar method to that
used for the \bxulnu\ modes. The decay channels of the charmed mesons
used are $D^+ \to K^- \pi^+ \pi^+$, $D^0 \to K^- \pi^+$,
$D^{*+} \to D^0 \pi^+$, $D^{*+} \to D^+ \pi^0$,
$D^{*0} \to D^0 \pi^0$ and $D^{*0} \to D^0 \gamma$.
The mass windows used in selecting the charmed meson are set to be
$[1.67, 2.07]\ \GeV\ $ for $D^+$,  
$[1.84, 1.89]\ \GeV\ $ for $D^0$, 
$[1.95, 2.25]\ \GeV\ $ for $D^{*+}$ and
$[1.99, 2.03]\ \GeV\ $ for $D^{*0}$.
Particle identification used is the same as for the \bxulnu\ modes.
The range of \qsq\ for charmed semileptonic decays is more narrow than for
the charmless case, and we break the range into the following three bins of
\qsq: 
$0 < \qsq < 3\ \GeVGeV$, $3 < \qsq < 6\ \GeVGeV$ and
$6 < \qsq < 16\ \GeVGeV$.

In performing the fits to the charm modes, the crossfeed components
used depend on the signal channel. For the \bdplnu\ channel the
background is divided into a \bdsplnu\ component and all other
backgrounds combined; for the \bdzlnu\ channel the crossfeed component
is \bdszlnu; for \bdsplnu\ the crossfeed component is \bdplnu; and for 
\bdszlnu\ the crossfeed component is \bdzlnu.
Following the same method employed for the charmless signal modes, the
partial branching fractions are extracted in each \qsq\ bin and summed
to give the total branching fraction. The results are shown in
Table~\ref{table:charm_bfsfitsum}, where they are compared to the
current PDG values~\cite{PDG}. The agreement is excellent.

\begin{table}[htp]
    \caption{\label{table:charm_bfsfitsum} Branching fractions for
      charm modes obtained from fitting in 3 \qsq\ bins and then summing
      the partial branching fractions, compared to Particle Data Group
      values. Errors in the ``Fitted'' column are statistical only, but
      take into account the finite MC statistics.}
    \begin{center}
      \begin{tabular}
      {@{\hspace{0.3cm}}c@{\hspace{0.3cm}}  
       @{\hspace{0.3cm}}c@{\hspace{0.3cm}}  
       @{\hspace{0.3cm}}c@{\hspace{0.3cm}}}
      \hline \hline
      Mode &
      Fitted   $\mathcal{B} \left[ \% \right]$ &
      PDG      $\mathcal{B} \left[ \% \right]$ \\
      \hline
      \bdplnu &
      \DpResultQsqSumFit &
      \DpResultPDG      \\
      \bdzlnu &
      \DzResultQsqSumFit &
      \DzResultPDG      \\
      \bdsplnu &
      \DspResultQsqSumFit &
      \DspResultPDG      \\
      \bdszlnu &
      \DszResultQsqSumFit &
      \DszResultPDG      \\
      \hline \hline
    \end{tabular}
  \end{center}
\end{table}


\subsection{Determination of $|V_{ub}|$}

The CKM matrix parameter $|V_{ub}|$ may be extracted from the partial
branching fraction $\Delta \mathcal{B}$ for \bpiplnu\ decay using the formula
$|V_{ub}| = \sqrt{\Delta \mathcal{B}/ \left( \tau_{B^0} \Delta \zeta \right)}$,
where $\tau_{B^0} = \left( 1.530 \pm 0.009 \right) \mathrm{ps}$ \cite{PDG} and
$\Delta \zeta = \Delta \Gamma / |V_{ub}|^2$ is the normalised partial
decay rate predicted from theoretical form factor calculations.

For $\Delta \zeta$ we take the values predicted from two approaches,
a LCSR calculation appropriate to the
kinematic region $\qsq < 16\ \GeVGeV\ $ \cite{ball-and-zwicky-05}
and recent Lattice QCD calculations valid for $\qsq > 16\ \GeVGeV\ $
\cite{hpqcd-06, fnal-04}.

We can improve the precision of our determination of $|V_{ub}|$ by
assuming that isospin symmetry is valid and combining the measurements of
the charged and neutral pion modes. To achieve this, the branching
fraction for the neutral pion mode in each bin of \qsq\ is first
multiplied by a factor $2 * \tau_{B^0}/\tau_{B^+}$ to account for the
difference in charged and neutral $B$ meson lifetimes, using
$\tau_{B^+}/\tau_{B^0} = 1.071 \pm 0.009$ \cite{PDG}. A weighted
average of this corrected \bpizlnu\ branching fraction and the
\bpiplnu\ branching fraction is then formed, using weights
$1/\sigma_i^2$ where $\sigma_i$ is the statistical error on the
partial branching fraction $i = \pi^+, \pi^0$. The systematic error is
taken as the weighted average of the individual systematic errors,
using the same weights. The resultant branching fraction obtained over
the full \qsq\ region is 
\begin{itemize}
\item[] $\mathcal{B}\left( \bpilnu \right)_{\mathrm{all\ }\qsq}
  = \left( \piresultqsqsumcomb \pm \pisyserrqsqsumcomb \right)
  \times 10^{-4}$
  \end{itemize}
and the partial branching fractions obtained in the low and high
\qsq\ ranges relevant to
$|V_{ub}|$ determination are
\begin{itemize}
\item[] $\Delta \mathcal{B}\left( \bpilnu \right)_{\qsq < 16\ GeV^2}
  = \left( \piresultqsqabcomb \pm \pisyserrqsqabcomb \right)
  \times 10^{-4}$
\item[] $\Delta \mathcal{B}\left( \bpilnu \right)_{\qsq > 16\ GeV^2}
  = \left( \piresultqsqccomb \pm \pisyserrqsqccomb \right)
  \times 10^{-4}$
\end{itemize}
respectively.

The results of the $|V_{ub}|$ determination are displayed in
Table~\ref{table:vub}, both for the case when the \bpiplnu\ partial
branching fractions are used, and when the above combined \bpiplnu\
and \bpizlnu\ partial branching fractions are used.
\begin{table}[htb]
    \caption{\label{table:vub} Values of $|V_{ub}|$ extracted from the
      measured \bpiplnu\ partial branching fractions and from the
      combined \bpiplnu\ and \bpizlnu\ partial branching fractions.
      The first error
      is statistical, the second systematic, and the third due to the
      theoretical errors quoted for the form-factor calculations.}
  \begin{center}
    \begin{tabular}
      {@{\hspace{0.35cm}}l@{\hspace{0.35cm}}  
       @{\hspace{0.35cm}}c@{\hspace{0.35cm}}  
       @{\hspace{0.35cm}}c@{\hspace{0.35cm}}  
       @{\hspace{0.35cm}}c@{\hspace{0.35cm}}  
       @{\hspace{0.35cm}}c@{\hspace{0.35cm}}}  
      \hline \hline
      & Mode & $\qsq\ [\GeVGeV$] & $\Delta \zeta\ [\textrm{ps}^{-1}]$
      & $|V_{ub}|\ [10^{-3}]$ \\
      \hline 
      Ball-Zwicky \cite{ball-and-zwicky-05} & $\pi^+$ &
      $< 16$ & $5.44 \pm 1.43$ &
      $3.2 \pm 0.3 \pm 0.1 ^{+0.5}_{-0.4}$ \\
      Gulez et. al. \cite{hpqcd-06}         & $\pi^+$ &
      $> 16$ & $2.07 \pm 0.57$ &
      $2.9 \pm 0.5 \pm 0.1 ^{+0.5}_{-0.3}$ \\
      Okamoto et. al. \cite{fnal-04}        & $\pi^+$ &
      $> 16$ & $1.83 \pm 0.50$ &
      $3.0 \pm 0.5 \pm 0.1 ^{+0.5}_{-0.3}$ \\
      \hline
      Ball-Zwicky \cite{ball-and-zwicky-05} & $\pi^+ + \pi^0$ &
      $< 16$ & $5.44 \pm 1.43$ &
      $3.1 \pm 0.2 \pm 0.1 ^{+0.5}_{-0.3}$ \\
      Gulez et. al. \cite{hpqcd-06}         & $\pi^+ + \pi^0$ &
      $> 16$ & $2.07 \pm 0.57$ &
      $3.1 \pm 0.3 \pm 0.1 ^{+0.6}_{-0.4}$ \\
      Okamoto et. al. \cite{fnal-04}        & $\pi^+ + \pi^0$ &
      $> 16$ & $1.83 \pm 0.50$ &
      $3.3 \pm 0.4 \pm 0.1 ^{+0.6}_{-0.4}$ \\
      \hline \hline
    \end{tabular}
\end{center}
\end{table}
  

\section{Summary}

In summary, we have made a study of the partial branching
fractions as a function of \qsq\ for five semileptonic decay channels
of $B$ mesons to charmless final states, using a full reconstruction
tag method. Summed over the three \qsq\ bins we obtain the following
estimates of the branching fractions: \
$\mathcal{B}\left( \bpiplnu \right)
  = \left( \pipshortresultqsqsumfit \right) \times 10^{-4}$,\
  $\mathcal{B}\left( \bpizlnu \right)
  = \left( \pizshortresultqsqsumfit \right) \times 10^{-4}$,\
  $\mathcal{B}\left( \brhoplnu \right)
  = \left( \rhopshortresultqsqsumfit \right) \times 10^{-4}$,\
  $\mathcal{B}\left( \brhozlnu \right)
  = \left( \rhozshortresultqsqsumfit \right) \times 10^{-4}$,\
  $\mathcal{B}\left( \bomegalnu \right)
  = \left( \omegashortresultqsqsumfit \right) \times 10^{-4}$,
  where the first error is statistical and the second systematic.
  From these branching fractions and theoretical form factor
  calculations, values for $|V_{ub}|$ have been obtained.
  All results are preliminary.
  
  Whilst the statistical precision of these measurements is limited at
  present, the potential power of the full reconstruction tagging
  method, when it can be used with larger accumulated $B$-factory data
  samples in the future, can clearly be seen. 

  \vspace{0.3cm}

  \acknowledgments
    
We thank the KEKB group for the excellent operation of the
accelerator, the KEK cryogenics group for the efficient
operation of the solenoid, and the KEK computer group and
the National Institute of Informatics for valuable computing
and SINET3 network support. We acknowledge support from
the Ministry of Education, Culture, Sports, Science, and
Technology of Japan and the Japan Society for the Promotion
of Science; the Australian Research Council and the
Australian Department of Education, Science and Training;
the National Natural Science Foundation of China under
contract No.~10575109 and 10775142; the Department of
Science and Technology of India; 
the BK21 program of the Ministry of Education of Korea, 
the CHEP SRC program and Basic Research program 
(grant No.~R01-2005-000-10089-0) of the Korea Science and
Engineering Foundation, and the Pure Basic Research Group 
program of the Korea Research Foundation; 
the Polish State Committee for Scientific Research; 
the Ministry of Education and Science of the Russian
Federation and the Russian Federal Agency for Atomic Energy;
the Slovenian Research Agency;  the Swiss
National Science Foundation; the National Science Council
and the Ministry of Education of Taiwan; and the U.S.\
Department of Energy.


\begin{thebibliography}{99}
\bibitem{CC}
Throughout this paper, 
the inclusion of the charge conjugate mode decay is implied
unless otherwise stated.
\end{thebibliography}
\end{document}